\documentclass[lettersize,journal]{IEEEtran}
\IEEEoverridecommandlockouts
\usepackage[usenames,dvipsnames]{xcolor}
\usepackage{hyperref}
\usepackage{caption}
\usepackage{subcaption}
\usepackage{cite}
\usepackage{amsmath,amssymb,amsfonts}
\usepackage{algorithmic}
\usepackage{graphicx}
\usepackage{textcomp}
\usepackage{xcolor}
\usepackage{multirow}

\def\BibTeX{{\rm B\kern-.05em{\sc i\kern-.025em b}\kern-.08em
    T\kern-.1667em\lower.7ex\hbox{E}\kern-.125emX}}
    
\begin{document}

\title{Adversarial Synthesis based Data augmentation for Code-Switched Spoken Language Identification
}

\author{\IEEEauthorblockN{Parth Shastri\IEEEauthorrefmark{1},
Chirag Patil\IEEEauthorrefmark{2}, Poorval Wanere\IEEEauthorrefmark{3}
Dr. Shrinivas Mahajan\IEEEauthorrefmark{4},
Dr. Abhishek Bhatt\IEEEauthorrefmark{5}, and Hardik Sailor\IEEEauthorrefmark{6}\thanks{\IEEEauthorrefmark{1},\IEEEauthorrefmark{2} Equal Contribution.}}

\IEEEauthorblockA{Dept. of Electronics and telecommunincation,
College of Engineering, Pune.\\
\IEEEauthorrefmark{6}Samsung Research and Development,
Bangalore.\\
Email: \IEEEauthorrefmark{1}shastripp18.extc@coep.ac.in,
\IEEEauthorrefmark{2}patilcn18.extc@coep.ac.in,
\IEEEauthorrefmark{3}wanereps18.extc@coep.ac.in,
\IEEEauthorrefmark{4}spm.extc@coep.ac.in
\IEEEauthorrefmark{5}bhatta.extc@coep.ac.in
\IEEEauthorrefmark{6}h.sailor@samsung.com}
}



\maketitle
\begin{abstract}
Spoken Language Identification (LID) is a critical subtask of Automatic Speech Recognition (ASR) that is used to categorise languages(s) in an audio segment. Automatic LID plays an important role in multilingual countries. In various countries, identifying a language becomes hard, due to the multilingual scenario where two or more than two languages are mixed together during conversation. Such phenomenon of speech is called as code-mixing or code-switching. This nature is followed not only in India but also in many Asian countries. Such code-mixed data is hard to find, which further reduces the capabilities of the spoken LID. Hence, this work primarily addresses this problem using data augmentation as a solution on the  data scarcity of the code-switched class. This study focuses on Indic language code-mixed with English. Spoken LID is performed on Hindi, code-mixed with English. This research proposes Generative Adversarial Network (GAN) based data augmentation technique performed using Mel spectrograms for audio data. GANs have already been proven to be accurate in representing the real data distribution in the image domain. Proposed research exploits these capabilities of GANs in speech domains such as speech classification, automatic speech recognition,etc. GANs are trained to generate Mel spectrograms of the minority code-mixed class which are then used to augment data for the classifier. Utilizing GANs give an overall improvement on Unweighted Average Recall by an amount of 3.5\% as compared to a Convolutional Recurrent Neural Network (CRNN) classifier used as the baseline reference.
\end{abstract}


\section{Introduction}
\IEEEPARstart{T}{he} language of a spoken utterance conveys a lot of information about the audio segment , which can be used later in more complex tasks like automatic speech recognition (ASR). Identification of a language of speech utterance in modern-day speech analysis has become a thrust area of research. Code-switching is an important aspect to study in articulations, especially in the Asian and African zones as it is observed in those nations in abundance. Identifying Code-Switching in an utterance by the means of classification is a supervised learning task that maps high-level audio features to labels or scalar values or probability scores of distinct classes. In this case, we need labeled dataset to build this classification system. This work studies the under-representation of the Code-Switched class and proposes Generative Adversarial Networks (GANs)\cite{b1} to generate the under-represented samples of the Code-Switched class.

The primary motive of this study is to use representation learning for synthesizing new code-switched data from input data distribution which can be used for the purpose of data augmentation. This data will be times-series code-switched Hindi-English speech signals which will be converted to image-like Mel spectrograms for the proposed GAN to learn the representation of the data. Recent approaches have used Convolutional Neural Networks with Mel spectrograms and cepstral features to classify audio\cite{https://doi.org/10.48550/arxiv.1609.09430, b2, vox}. A combination of RNNs with CNNs can learn both the spatial and temporal signal information and prove to be more effective than using either one of them separately\cite{b3}.

Indic code-switched data is hard to find in the wild and needs a lot of preprocessing to meet desired criteria for a task. Generally, in order to fulfill the data requirements, data augmentation is used. Rangan et al.(2020) have produced results on a similar task using spectral augmentation techniques\cite{b26}.
The existing audio data augmentation techniques work well with classifiers but do not reflect the ground truth of variances in speech. For instance, a common form of data augmentation is to speed up or slow down existing speech; while this technique might improve accuracy, it does not reflect the true nature of human speech\cite{b4}. A synthesis-based augmentation technique, on the other hand, is more likely to reflect changes in rate more accurately. Therefore, it is believed that synthesis-based augmentation should yield even better results. Hence, this proposed approach focuses on the task of generating synthetic code-switched speech signals and further increase the already existing dataset of speech signals.
\IEEEpubidadjcol

Generative Adversarial Networks\cite{b1} have proven to be effective in learning image representations. GANs have been used to generate realistic images in complex tasks \cite{b28, b29, cyclegan}. Recently GANs have proven their ability to generate realistic in-distribution samples and are used in augmentation tasks. In \cite{b5}\cite{b30}, GANs are used to augment data and address data imbalance. Treating Mel spectrograms as images, we try to use the strong point of GANs to generate in-distribution code-switched samples.
In this work, focus on the Deep Convolutional GAN (DCGAN) architecture \cite{b6}, since the purpose is not to generate high-quality speech signals. Instead, our investigation concerns generating acceptable speech signals in the form of spectrograms which can be used further for a classifier as additional data to train it. A Convolutional Recurrent Neural Network similar to \cite{b3}, is built to classify the synthesized images and original corpus as a larger augmented dataset and compare it to the original baseline corpus.
This study represents novelty in using GANs to augment code-switched samples.

The further sections are organized as follows: An overview of GANs is presented in ''Sec.~\ref{sec:overview}. ''Sec.~\ref{sec: Proposed method}'' discusses the proposed method architecture and the motivation for the use of GANs. It also showcases the results obtained by training the GANs on the minority class. ''Sec.~\ref{sec: experiments}'' provides details about the dataset and establishes the experimental setup of the baseline methods and the proposed method of GAN-based augmentation for spoken LID. The results of the experiments and their detailed comparison are done in ''Sec.~\ref{sec: performance}''. The final conclusions and the future scope is discussed in ''Sec.~\ref{sec: conclusion}''. 
\section{GANs: an Overview}\label{sec:overview}
Generative Adversarial Networks (GANs) are generative networks that work on the principle of a zero-sum non-cooperative min-max game between two players. In this game, the Generator (G) tries to learn the data distribution $P_r$ via the mistakes of the Discriminator (D). In general, GANs try to minimize the divergence between the model distribution $P_g$ and the data distribution $P_r$. The G and D are trained simultaneously, where D tries to differentiate between the real input and a synthetic input generated by the generator. As the training proceeds, the generator tries to generate more realistic input so as to fool the discriminator. The objective of the network is given in equation \ref{eq:loss}.
\begin{equation}\label{eq:loss}
\begin{aligned}
\mathcal{L_{\text{GAN}}} = \min_{G}\max_{D}\mathbb{E}_{x\sim P_{\text{r}}(x)}[\log{D(x)}] \\ 
+ \mathbb{E}_{z\sim P_{\text{z}}(z)}[1 - \log{D(G(z))}]
\end{aligned}
\end{equation}

Where $x$ is the sample from the data distribution $P_r$ and $z$ is the input to the generator, which can be a noise vector or can be a conditioning vector as in the Conditional GANs framework \cite{b14}, taken from a distribution of such vectors $P_z$. Here D and G can be any differentiable functions representing the discriminator and the generator, respectively.

While traditional GANs have been shown to produce realistic outputs, they are subject to high instability in training and problems like mode collapse and vanishing gradient during training. To overcome these issues while training in GANs, Wasserstein-GAN (WGAN) was proposed in \cite{b15}. This method mitigates GAN training issues by minimizing the Earth-Mover (EM) distance between the real and the generated distribution. The loss function of WGAN is showcased in equation \ref{eq:WGAN loss}. In this method, the discriminator network is replaced by a network called critic, also denoted by D.
\begin{equation}\label{eq:WGAN loss}
    \mathcal{L_{\text{WGAN}}} = \min_{G}\max_{||D||_L\leq1}\mathbb{E}_{x\sim P_{\text{r}}(x)}[D(x)] - \mathbb{E}_{z\sim P_{\text{z}}(z)}[D(G(z)]
\end{equation}

Where D is required to be 1-Lipschitz, this constraint can be enforced by weight-clipping as done in the original paper\cite{b15}, but as discussed in the original paper, it is a terrible way to enforce the constraint. To address this, a new method to enforce this constraint was proposed by  Gulrajani et al. in WGAN-GP \cite{b16}. They propose a Gradient-Penalty that is added directly to the loss function of the WGAN. This penalty is shown in equation \ref{eq:gp} and \ref{eq:x}.
\begin{equation}\label{eq:gp}
    GP = \mathbb{E}_{\hat{x}\sim P_{\hat{x}}}[(||\nabla_{\hat{x}}D(\hat{x})||_2 - 1)^2]
\end{equation}

Where,
\begin{equation}\label{eq:x}
\begin{aligned}
    \hat{x} = \alpha\Tilde{x} + (1 - \alpha)x\\
    \text{with, } \alpha \sim U(0, 1)
\end{aligned}
\end{equation}

The total loss of the WGAN-GP is given by the equation \ref{eq:wgangp loss}.
\begin{equation}\label{eq:wgangp loss}
    \mathcal{L_{\text{WGAN-GP}}} = \mathcal{L_{\text{WGAN}}} + \lambda_{\text{gp}}GP
\end{equation}
This method uses a conditional architecture, which generates a distribution conditioned on the conditional vector described in ''Sec.~\ref{sec: Proposed method}'' and follows the training algorithm mentioned in the original paper\cite{b17} with the same hyperparameters.

\section{Proposed Method}\label{sec: Proposed method}

\subsection{Motivation}\label{motivation}

In the real-world, Code-switched datasets are hard to find, which leaves us with fewer examples in hand to perform classification on. However, very few works have focused on this issue. Many works focus on the Chinese languages leaving very less work on Indic datasets\cite{b7}. Many works either use signal-based transformations\cite{b8}, \cite{b26} or the popular SpecAugment\cite{b9}, which uses time-warping, time-masking, and frequency-masking to produce deformations in the logarithmic Mel spectrograms. We use SpecAugment for comparison with the proposed GAN method.

The purpose of this study is to validate the method of GAN-based representation learning to augment a specific class of a dataset. In the absence of adequate training data, data augmentation is done in order to increase the effective size of the dataset. This method has proven highly effective in image classification. This method has an effect of making data effectively larger by feeding the model multiple augmented versions of the same data, over the course of training. Another method includes applying transformations to the data and adding that data to the dataset physically, but this requires additional memory. This paper uses the former method and uses a GAN to do so.

\begin{figure}[h]
    \centering
    \includegraphics[width=\linewidth]{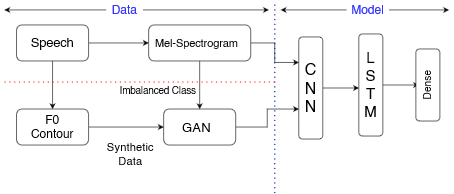}
    \caption{Proposed methodology for spoken LID using CNN-LSTM Network.}
    \label{fig:proposed_method}
\end{figure}

\subsection{Method Description}
As shown in Fig. \ref{fig:proposed_method}, the CRNN-based spoken LID system uses GANs to augment the speech signals. This system receives the 2-D Mel spectrograms as the input. The CNN layer captures the spatial invariance and the Bi-LSTM layer captures the sequential temporal context from the input. Only the under-represented code-mixed class is passed through the GAN for augmentation. The final Dense (Fully-connected) layer maps the inputs to classwise probabilities between 0 and 1.

\begin{figure}[t]
    \centering
    \begin{subfigure}[b]{0.45\textwidth}
        \centering
        \includegraphics[width=\textwidth]{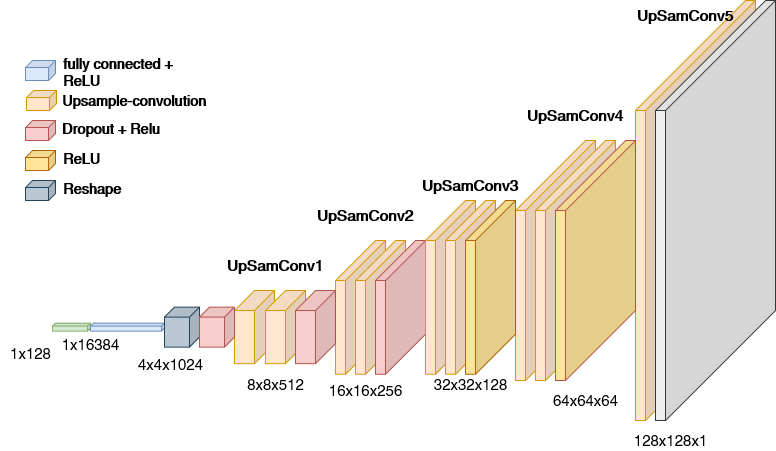}
        \caption{Generator (G)}
        \label{fig: G arch}
    \end{subfigure}
    \begin{subfigure}[b]{0.5\textwidth}
        \centering
        \includegraphics[width=\textwidth]{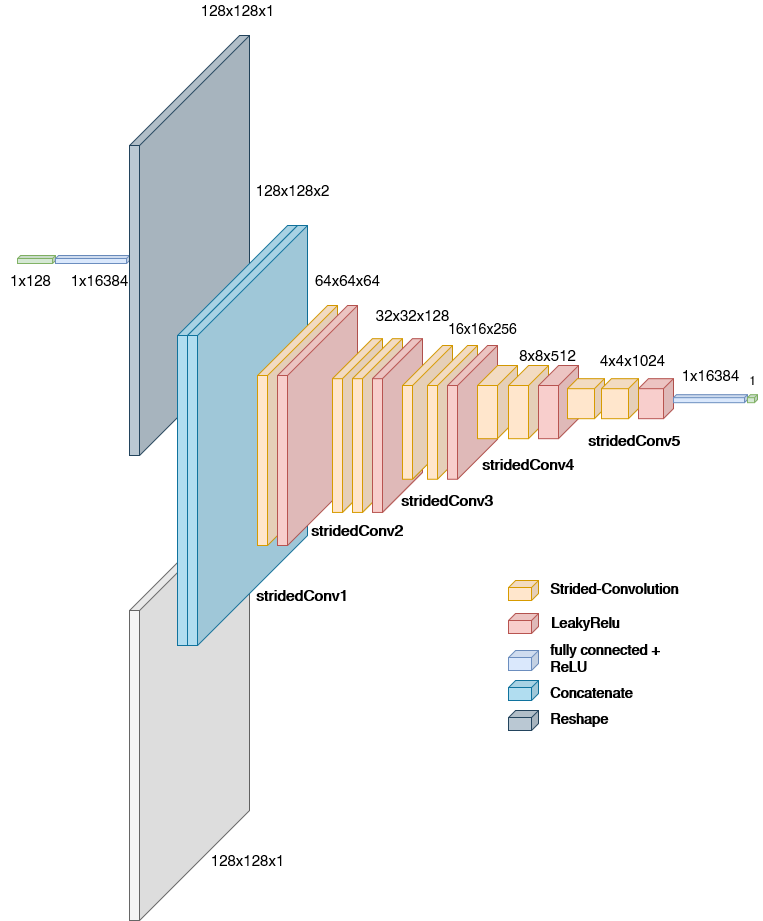}
        \caption{Critic (D)}
        \label{fig: D Arch}
    \end{subfigure}
    \caption{Architecture of the proposed GAN}
    \label{fig1}
\end{figure}

This work uses a DCGAN-based Conditional architecture\cite{cgan}, conditioned on the fundamental frequency similar to \cite{b10}.The fundamental frequency represents the pitch information of a speaker and provides for a good conditioning factor. The $\log{F0}$ contour is extracted using the WORLD algorithm \cite{b11}. The DCGAN has already proved its potential for generating images, we see that image like spectrograms can also be generated using GANs\cite{b6}.

The GAN is trained only on the code-Switched data, to learn the distribution of the code-switched class only. The generator contains a series of upsampling convolutions and the discriminator uses strided convolutions, the architecture is similar to SpecGAN with some changes\cite{b12}. However, the original SpecGAN was used to generate log-scale spectrograms directly, the proposed work leverages the architecture for Mel- spectrograms. The Mel-scale is a powerful tool in the world of speech processing as it represents the way humans perceive speech in the real-world. Mel spectrograms are adapted in many speech-classification an speech-recognition tasks \cite{b13}. 

The proposed architecture is illustrated in Fig. \ref{fig1}. The Generator (G) illustrated in Fig. \ref{fig: G arch} and the Critic (D) illustrated in Fig. \ref{fig: D Arch} play an adversarial game to generate realistic Mel spectrograms from the conditioned $\log{F0}$ vector.
The Dense layer is used to project the input conditioning vector to a higher dimension followed by the use of five upsampling convolutional layers. The final convolutional layer of G is used to provide the spectrogram image as output. Further, the conditioning vector and the spectrogram image has been given to D as inputs. The conditioning vector is first passed through a fully-connected network similar to the CGANs\cite{cgan} which in this case contains a Dense layer which projects the input to a higher dimension and later is reshaped to a shape similar to the spectrogram image. Both the tensors are then concatenated along the channel axis and later fed to the convolutional layers. Five convolutional layers followed by a dense layer used to classify the real and fake spectrograms are further present in D. The detailed architecture is depicted in tabular form in "Table. ~\ref{tab:Generator_table}" and "Table. ~\ref{tab:Discriminator_table}"
\begin{table}[!h]
    \centering
    \caption{Critic architecture.}
    \begin{tabular}{|c|c|c|}
        \hline
        \multirow{2}{*}{\textbf{OPERATION}}& \multirow{2}{*}{\textbf{KERNEL SIZE}} &\multirow{2}{*}{\textbf{OUTPUT SIZE}} \\
        & & \\
        \hline
        InputLayer1 $F0$ & - &  [(n,128)]\\
        Dense& (128,16384) &  (n,16384)\\
        Reshape & - &  (128,128,1)\\
        InputLayer2 (spectrogram) & - &  (128,128,1)\\
        \cline{1-2}
        \multicolumn{2}{|c|}{\multirow{2}{*}{Concatenate (Reshape,  InputLayer2)}}  &  (128,128,2)\\
         \multicolumn{2}{|c|}{} & \\
        \cline{1-2}
        Conv2D & (25,2,64) & (64,64,64)\\
        LeakyReLU & - & (64,64,64)\\
        Conv2D & (25,64,128) & (32,32,128)\\
        LeakyReLU & - & (32,32,128)\\
        Conv2D & (25,128,256) & (16,16,256)\\
        LeakyReLU & - & (16,16,256)\\
        Conv2D & (25,256,512) & (8,8,512)\\
        LeakyReLU & - & (8,8,512)\\
        Conv2D & (25,512,1024) & (4,4,1024)\\
        LeakyReLU & - & (4,4,1024)\\
        Reshape & - &  (n,16384)\\
        Dense & (16384,1) & (n,1)\\
        \hline
        \multicolumn{2}{|c|}{\textbf{Total number of parameters}} & \textbf{1.9M}\\
        \hline
    \end{tabular}
    \label{tab:Generator_table}
\end{table}

\begin{table}[!h]
    \centering
    \caption{Generator Architecture.}
    \begin{tabular}{|c|c|c|}
        \hline
          \multirow{2}{*}{\textbf{OPERATION}}& \multirow{2}{*}{\textbf{KERNEL SIZE}} &\multirow{2}{*}{\textbf{OUTPUT SIZE}} \\
        & & \\
         \hline
         Input $F0$& - &  (n,128)\\
         Dense& (128,16384) &  (n,16384)\\
         Reshape & - &  (n,4,4,1024)\\
         Dropout& - &  (n,4,4,1024)\\
         ReLU&  & - (n,4,4,1024)\\
         UpsampleConv& (25,1024,512) &  (n,8,8,512)\\
         Dropout& - &  (n,8,8,512)\\
         ReLU& - &  (n,8,8,512)\\
         UpsampleConv& (25,512,256) &  (n,16,16,256)\\
         ReLU& - &  (n,16,16,256)\\
         UpsampleConv& (25,256,128) &  (n,32,32,128)\\
         ReLU& - &  (n,32,32,128)\\
         UpsampleConv& (25,128,64) &  (n,64,64,64)\\
         ReLU& - &  (n,64,64,64)\\
         UpsampleConv& (25,64,1) &  (n,128,128,1)\\
         \hline
         \multicolumn{2}{|c|}{\textbf{Total number of parameters}} & \textbf{1.9M}\\
         \hline
    \end{tabular}
    
    \label{tab:Discriminator_table}
\end{table}

The GAN training is highly unstable and many works have tried to mitigate the problems caused during the training of GANs\cite{b17}\cite{b15}\cite{b18}\cite{nips}. The best improvement in stability was showcased by using the Wasserstein distance as the loss along with Gradient Penalty which enforces a Lipschitz constraint by use of a penalty in the loss of the critic \cite{b16}. Many state-of-the-art GANs have adapted this loss and shown it to be stable.

This work optimizes the Wasserstein loss coupled with the gradient penalty to train the G and the D. The input conditioning vector is directly fed to the G in which noise is added through the use of fixed Dropout layers in the initial layers similar to the pix2pix generator\cite{b19}. The fake images generated by the G are then fed to D in batches separately along with the real spectrogram batches. Both the G and the D networks are then trained in an adversarial manner to optimize their parameters through the Wasserstein objective function. We found it useful to guide the training of the network by further adding a reconstruction loss, depicted in equations \ref{eq:recon}, \ref{eq:added recon}. This type of loss is used in conditional GANs like \cite{b19}, \cite{cyclegan}.
\begin{equation}\label{eq:recon}
    \mathcal{L_{\text{recon}}} = \mathbb{E}_{x,z}[||x - G(z)||_1]
\end{equation}
\begin{equation}\label{eq:added recon}
    \mathcal{L_{\text{total}}} = \mathcal{L_{\text{WGAN-GP}}} + \lambda_{\text{recon}}\mathcal{L_{\text{recon}}}
\end{equation}
Where $\lambda_{\text{recon}}$ is a hyperparameter. This paper uses $\lambda_{\text{recon}} = 10$.
\subsection{Details of implementation}\label{impldet}
The spectrograms are scaled in range [-1, 1] with the help of min-max normalizer similar to \cite{b20}.The spectrograms are obtained by using a FFT window of size 1024 and an overlap of 16ms. The $\log{F0}$ contour is obtained using similar parameters to get a vector of length 128. The final layer of the G uses \emph{tanh} activation due to the range of the spectrogram images. The spectrograms obtained are of size [128, 128]. In the generator batch-normalization is used while layer-normalization is used in the  critic. According to \cite{b21} this combination of normalization schemes proves to be effective in GAN training. The D uses LeakyRelu similar to DCGAN\cite{b6}. The Dropout layers in G have a $p=0.5$ and the LeakyRelu in the D has $alpha=0.2$. In addition, Adam optimizer is used with a learning rate of $1\times10\textsuperscript{-4}$ and $\beta_1=0.5,\beta_2=0.9$. The F0 contour is first converted into semitone scale, to explore the perceptually relevant information \cite{b22} given by the equation \ref{eq:semitone}:
\begin{equation}\label{eq:semitone}
    ST=39.87\times\log{\frac{F0}{50}}
\end{equation}
This $\log{F0}$ is later normalized in the range [0, 1] using min-max scaler. The network is trained for 150k iterations with a batch size of 8, but converges around 115k iterations.
\subsection{Architecture Details}
The proposed GAN architecture is similar to SpecGAN\cite{b12} with some architectural changes it the network. The changes are as followed:(a) This work uses a CGAN architecture, using a conditioning information vector in the form of $\log{F0}$ contour.
(b) The D uses layer-normalization as opposed to no normalization in the SpecGAN critic.
(c) Instead of feeding a noise latent vector $Z$ directly to the G it is implicitly provided through the Dropout in the initial layers similar to pix2pix\cite{b19}.
(d) The G uses batch-normalization and upsampling convolutions instead of transposed Convolutions.
(e) In both G and D phase shuffle is not used.

\subsection{Dataset for training the GAN}
MUCS ASR Challenge\footnote{The data was taken from openslr \url{https://www.openslr.org/104/}} dataset \cite{b23} was taken to obtain the Hindi-English code-mixed recordings, the dataset contains Hindi-English Code-mixed lecture recordings in \emph{".wav"} format, encoded using 16-bit PCM and having a sample rate of 16kHz. The data contains a segments file which stores the time-stamps of the code-mixed speech segments. The train-data after cleaning and trimming the silences contains $\sim23$ hours of speech audio. The test data contains $\sim4$ hours of speech audio. The train-data is used for training the GAN to generate artificial samples as close to the code-switched spectrograms as possible. The test data is used in the classification task later.

\subsection{Evaluation of GAN}
The proposed GAN is evaluated based on the Frechet Inception Distance (FID) metric. It was proposed by Heusel et al.. \cite{fid}. The FID takes into account the statistics of the real images and generated images and compares them. It is an improvement on the Inception Score (IS) proposed by \cite{impgans}. First, the features are extracted from the generated images and the real images from a pretrained InceptionV3 network \cite{inceptionv3}. As the InceptionV3 network is trained on Imagenet dataset \cite{imagenet}, the features obtained by passing spectrograms through this network will not be meaningful. In the WaveGAN paper, by Donahue et al.. a separate network is trained on the dataset used in that paper to calculate the IS. Similarly, the CRNN baseline proposed in "Sec.\ref{sec:baseline_method}'' is used to extract meaningful features.

\begin{figure}[htbp]
    \centering
    \includegraphics[width=\linewidth]{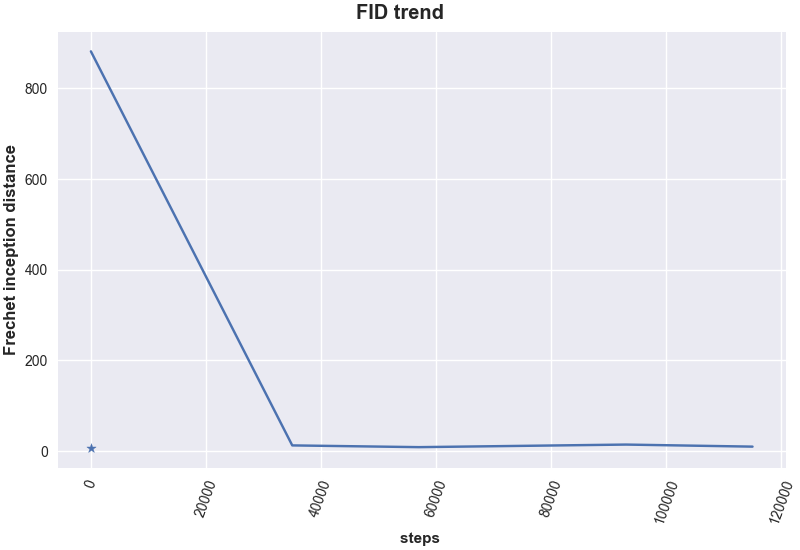}
    \caption{The FID trend for $n=2400$  with respect to the number of training iterations(steps) of the proposed GAN.}
    \label{fig:fid_trend}
\end{figure}

\begin{figure}[!h]
    \centering
    \begin{subfigure}[b]{0.5\textwidth}
        \centering
        \includegraphics[width=\textwidth]{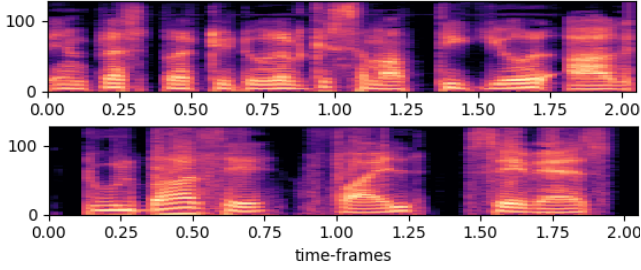}
        \caption{Real Mel spectrograms}
        \label{fig: real_spec}
    \end{subfigure}
    \hfill
    \begin{subfigure}[b]{0.5\textwidth}
        \centering
        \includegraphics[width=\textwidth]{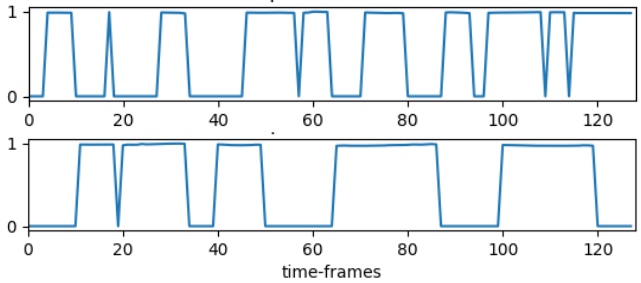}
        \caption{Normalized $\log{F0}$ contour}
        \label{fig: nr_f0}
    \end{subfigure}
    \begin{subfigure}[b]{0.5\textwidth}
        \centering
        \includegraphics[width=\textwidth]{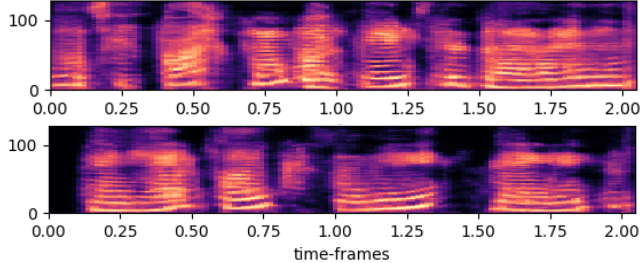}
        \caption{Generated Mel spectrograms}
        \label{fig: gen_spec}
    \end{subfigure}
    \caption{Generated images from the proposed GAN. (a) Real Mel spectrograms from the test data. (b) Normalized $\log{F0}$ contours of the audio associated with the real spectrograms. (c) Generated Mel spectrograms by the proposed method.}
    \label{fig:gen_examples}
    
\end{figure}

The penultimate convolution layer of the CRNN baseline is used to extract the features. The FID treats the real and generated samples as samples from two multivariate gaussian distributions. The mean ($\mu_{r}, \mu_{g}$) and covariance ($\Sigma_{r}, \Sigma_{g}$) of the features is then calculated. Finally, the FID is calculated by using equation \ref{eq:fid}. A lower score indicates good quality of generated samples.
\begin{equation}\label{eq:fid}
FID = ||\mu_{r} - \mu_{g}||^2 + \text{Tr}(\Sigma_{r} + \Sigma_{g} - 2(\Sigma_{r}\Sigma_{g})^\frac{1}{2})
\end{equation}
''Table.~\ref{tab:fid comparison}'' compares the FID values obtained by using various sample sizes, the FID score of the test data is also mentioned. ''Fig.~\ref{fig:fid_trend}'' shows the FID ($n=2400$) trend for different iterations of GAN training. It can be clearly observed that the FID decreases with increasing training iterations. In Fig \ref{fig:gen_examples} some examples of the spectrograms generated from the proposed architecture are displayed. The Real images are shown in "Fig.~\ref{fig: real_spec}" and the generated images by the proposed architecture are shown in "Fig.~\ref{fig: gen_spec}".
\begin{table}[h]
    \setlength{\tabcolsep}{10pt} 
    \renewcommand{\arraystretch}{1.5}
    \centering
    \caption{Comparison of the FID scores of the proposed GAN and the real test data, for different sample sizes of real and generated images. }
    \begin{tabular}{|c|c|}
        \hline
         \textbf{MODEL NAME (n=sample size)}&\textbf{FID}  \\
         \hline
         Real-test data & 6.15 \\
         \hline
         proposed GAN (n=500)& 11.14\\
         \hline
         proposed GAN (n=1000)& 10.29\\
         \hline
         proposed GAN (n=1000)& 9.30\\
         \hline
         proposed GAN (n=2000)& 9.47\\
         \hline
         proposed GAN (n=2400)& 9.60\\
         \hline
         \textbf{AVG. FID} (proposed GAN)& 9.97\\
         \hline
    \end{tabular}
    \label{tab:fid comparison}
\end{table}

\section{Experiments}\label{sec: experiments}
\subsection{Language identification dataset}\label{sec: dataset}
The data for the three-class classification \emph{"Hindi", "English" and "Hindi-English"} is not readily available. In this work we create a small dataset to train the classifier using some existing Language Identification datasets and taking audio from some existing ASR datasets. We used the datasets that contained Indian English to imitate the Indic landscape.The combination is done as follows:
\begin{itemize}
\item NPTEL Indian English dataset \cite{b24} is used to borrow the \emph{"English"} examples.
\item NISP dataset \cite{b25} is used to borrow some examples from the \emph{"English"} class.
\item The CVIT IndicTTS dataset \cite{indictts} is used to borrow the samples from the \emph{"Hindi"} class.
\item The test data from the MUCS ASR \cite{b23} Challenge data is used to borrow the code-mixed \emph{"Hindi-English"} examples.
\end{itemize}
The audio from these datasets are recorded under similar conditions making them good candidates for this combination.
The borrowed samples are converted to 16-bit PCM if they are not in given format and then resampled to 16KHz sampling rate.  The created dataset is then split 80\%-20\% to create a test dataset which is later used for evaluation. The dataset created is kept balanced initially with same number of examples for each class: namely \emph{english, hindi} and \emph{hindi-english}. The generated dataset after splitting and preprocessing contains $\sim6$ hours of audio. 

\subsection{Classification and feature extraction}\label{AA}
We use an architecture similar to CRNN architecture for the classification of examples. This architecture classifies three classes, viz. Hindi, English, Hindi-English. Thus it converts to a three-class spoken LID task, out of which one is Hindi code-mixed with English. The architecture of the CRNN classifier is given in fig. The classifier is fed Mel spectrograms of size [128,128], similar to the GAN training depicted in Sec \ref{impldet}. These Mel spectrograms are calculated from audio segments randomly cropped or padded to a fixed length of $\sim2$ secs. The overlap window of 16 ms is used during the extraction of spectrograms, and this gives $25\%$ overlap ratio, 128 Mel-filterbanks are used to produce 128 frequency bins. The log-scale spectrograms are calculated and are later converted into decibel (dB) scale and then are normalized in the range of [0, 1].
\subsection{Train-Test data Split}
The Hindi-English dataset is extracted from spoken   tutorials. These tutorials cover a range of technical topics and the code-switching predominantly arises from the technical content of the lectures. The segments file in the baseline recipe provides sentence time-stamps. These time-stamps were used to derive utterances from the audio file. 
We follow 80\%-20\% train validation split and 5-fold cross-validation or evaluation for selection of the best model. A separate data subset is kept for testing purposes, as already mentioned before. (refer to "Sec.~\ref{sec: dataset}").
\subsection{Dataset imbalance}
We simulate data set imbalance by randomly leaving out 80\% of samples from the Hindi-English code-switched class, similar to \cite{gansser}. This is performed in order to demonstrate real world conditions of the under-representation of this class. "Table.~\ref{tab:examples_per_class}" shows the number of examples per class after applying the dataset imbalance heuristic.
\begingroup
\setlength{\tabcolsep}{10pt} 
\renewcommand{\arraystretch}{1.5}
\begin{table}[htbp]
    \centering
    \caption{Number of examples per class.}
    \begin{tabular}{ccc}
    \hline
         \textbf{CLASS} & \textbf{TRAIN} & \textbf{TEST}  \\
         \hline
         English&2841&400\\
         
         Hindi&3070&362\\
         
         Hindi-English&627&400\\
         \hline
         \textbf{Total}&6538&1162 \\
         \hline
    \end{tabular}
    \label{tab:examples_per_class}
\end{table}
\endgroup

\subsection{Refernce baselines}\label{sec:baseline_method}
For comparison, we implement four baselines other than the proposed methodology. The first method is simply training the model on the under-represented dataset using a Convolutional Extractor for spatial feature extraction and then using an LSTM network to exploit the temporal information. This architecture is similar to CRNN \cite{b3}. Then, to augment the spectrograms from the code-mixed class, we use the SpecAugment method \cite{b9} as already discussed in "Sec.~\ref{motivation}". The method is used on the baseline that we defined. The SpecAugment parameters used are obtained by trial and error. Although, by carefully tuning the parameters, more performance can be achieved. Furthermore, traditional methods of time-stretching the audio and pitch-shift are used for the augmentation purpose to further compare the performance of the proposed system.
The details of the baselines are as follows:
\begin{itemize}
    \item \textbf{No-Augmentation} The CRNN classifier is trained on the imbalanced dataset without using any data augmentation.
    \item \textbf{SpecAugment} The classifier was trained by using the SpecAugment technique of time-masking and frequency-masking introduced by Google Brain. The parameters used were $F=13$ and $T=20$.
    \item \textbf{Time-stretch} The classifier was trained using the traditional method of randomly time-stretching the raw audio signal with a rate in the range $\textit{r} \in [0.5, 1.5)$.
    \item \textbf{Pitch-shift} The pitch of the audio signal is randomly shifted in the range of $n\_steps\in[-4, 4]$. This technique creates augmentation by shifting the pitch of the raw audio signal by semitones in the given range.
\end{itemize}

\subsection{Evaluation}\label{sec: eval}
The spoken LID system performance is evaluated using Accuracy, Precision, F1-Score, and Unweighted Average Recall (UAR). UAR and F1-score are widely used in the case of imbalanced classification and are more reliable than the accuracy in such scenarios. Equations \ref{acc}, \ref{prec}, \ref{recall}, \ref{f1}, \ref{uar} present the above-mentioned metrics in more detail.

\begin{equation}\label{acc}
    Accuracy = \frac{TP + TN}{TP + TN + FP + FN}
\end{equation}
\begin{equation}\label{prec}
    Precision = \frac{TP}{TP + FP} 
\end{equation}
\begin{equation}\label{recall}
    Recall = \frac{TP}{TP + FN}
\end{equation}
\begin{equation}\label{f1}
    F1-Score = \frac{2\cdot Precision\cdot Recall}{Precision + Recall}
\end{equation}

where \emph{TP, TN, FP, and FN} stand for the True Positives, True negatives, False Positives, and False Negatives, respectively. F1-score is just the harmonic mean of precision and recall.

\begin{equation}\label{uar}
    UAR = TPR\cdot \frac{p}{p+n} + FPR \cdot \frac{n}{p+n}
\end{equation}

where \emph{TPR and TNR} stand for the True Positive Rate and True Negative Rate respectively, and \emph{p} and \emph{n} mean the total no. of positives and the total no. of negatives.

\subsection{Training Method}
\begin{figure}[htbp]
    \centering
    \includegraphics[width=\linewidth]{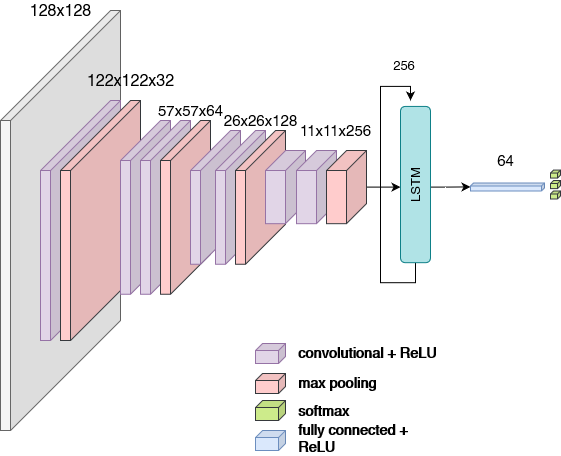}
    \caption{The classifier architecture.}
    \label{fig:classifier}
\end{figure}

A three class classifier model based on the architecture of \cite{b3} was trained on the dataset as already discussed in "Sec.~\ref{sec: dataset}". The architecture of the classifier is shown in "Fig.~\ref{fig:classifier}" The classifier was trained with Stochastic gradient descent with Adam optimizer. A learning rate of $1\times10\textsuperscript{-5}$ was used. The model was trained for approximately $\sim30$ epochs using early stopping to prevent overfitting and was evaluated on a balanced test dataset with the metrics as already mentioned (refer to "Sec.~\ref{sec: eval}").

\section{Performance and Results}\label{sec: performance}
\begin{figure}
    \centering
    \includegraphics[scale=0.45]{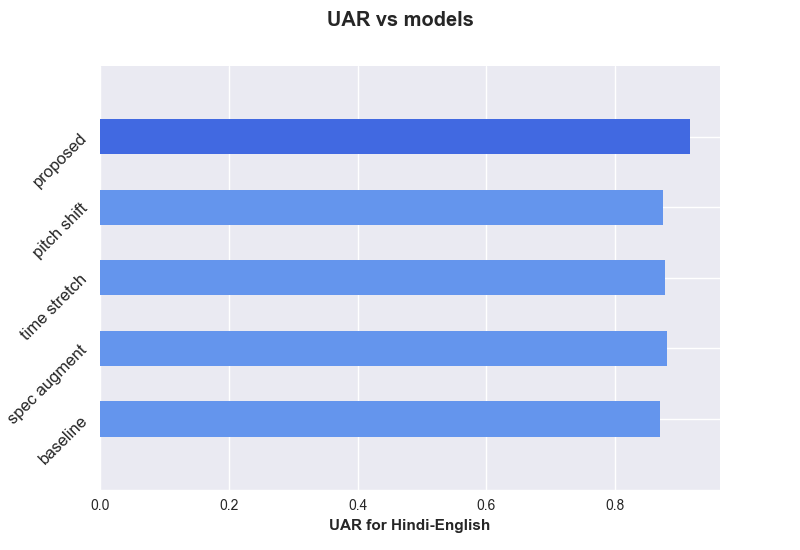}
    \caption{Comparison of UAR scores.}
    \label{fig:uar_comparison}
\end{figure}

\begin{figure}
    \centering
    \includegraphics[scale=0.4]{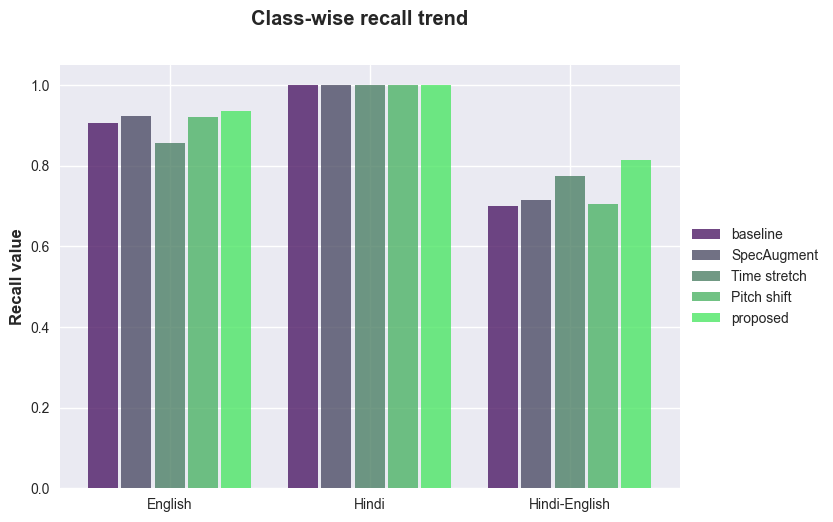}
    \caption{Comparison classwise recall scores.}
    \label{fig:cw_recall}
\end{figure}

\begin{figure}
    \centering
    \includegraphics[scale=0.4]{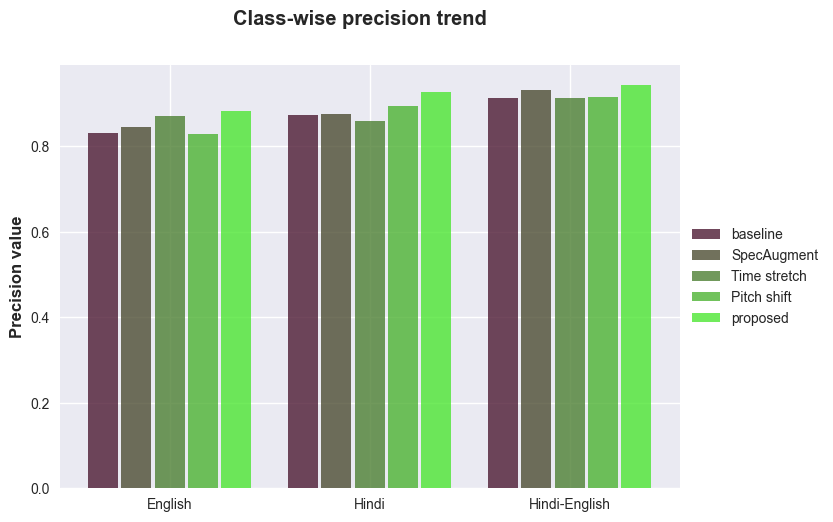}
    \caption{Comparison of classwise precision scores.}
    \label{fig:cw_prec}
\end{figure}

In "Table.~\ref{tab:acc_table}", we compare the accuracy of the proposed method with that of the baselines. The last column shows the percentage (\%) increase from the baseline.
In "Table.~\ref{tab:comparison_table}", we demonstrate the performance achieved using our dataset. The rows in the table correspond to the training methods described in "Section ~\ref{sec:baseline_method}". Each column represents the metric monitored. We see that the proposed method achieves an 8.1\% relative performance improvement compared to the baselines in the F1-score. The proposed method also achieves balanced precision and recall scores compared with the other methods. The time-stretching method provided an improvement in the UAR by 0.8\% and the F1-score of 4.4\%. The pitch shift method showed an improvement in the F1-Score by 0.02\%. In "Fig.~\ref{fig:uar_comparison}", the UAR metric is compared for each of the methods and the proposed model performs better than all the other methods and shows an improvement of 4.7\% over the baseline. The model performs better than the other models without hurting the scores of the other two classes, the comparison of class wise recall and class wise precision is done in ''Fig.~\ref{fig:cw_recall}'' and ''Fig.~\ref{fig:cw_prec}'' respectively.
\begingroup
    \setlength{\tabcolsep}{10pt}
    \renewcommand{\arraystretch}{1.5}
\begin{table}[htbp]
    \centering
    \caption{Accuracy comparison}
    \begin{tabular}{|c|c|c|}
    \hline
    \textbf{METHODS}& \textbf{(\%)ACCURACY}&\textbf{(\%)INC.} \\
    \hline
    \textbf{No-aug (baseline)}& 87.0  &-\\
    \hline
    \textbf{SpecAugment\cite{b9}}&  88.0 &1.0\\
    \hline
    \textbf{Time Stretch}&   87.8&0.8\\
    \hline
    \textbf{Pitch Shift}& 87.3  &0.3\\
    \hline
    \textbf{GAN (proposed)}&  \textbf{91.7} &\textbf{4.7}\\
    \hline
    \end{tabular}
    \label{tab:acc_table}
\end{table}
\endgroup
\begingroup
\renewcommand{\arraystretch}{1.5}
\begin{table}[!h]
\centering
\caption{Model performance on the minority class.(The values mentioned in the table are recorded from the augmented minority "Hindi-English" class.)(\emph{Precision, recall and F1-score})}
\begin{tabular}{|p{1.5cm}|p{1.5cm}|p{1.5cm}|p{1.5cm}|p{0.8cm}|}
 \hline
 \multicolumn{5}{|c|}{Comparison of metrics} \\
 \hline
 &\textbf{Precision} & \textbf{Recall}& \textbf{F1-Score}&\textbf{(\%)Inc.}\\
 \hline
 \textbf{No-aug (baseline)}& 0.9145 &0.7009& 0.7935&-\\
 \hline
 \textbf{SpecAugment \cite{b9}}&0.9331 &0.7151& 0.8097&1.6\\
 \hline
\textbf{Time stretch}&0.9128  & 0.7749 & 0.8382&4.4\\
\hline
\textbf{Pitch shift}&0.9148 &0.7037 &0.7955 &0.02\\
\hline
\textbf{GAN (proposed)}&0.9439 &0.8148 &\textbf{0.8746} &\textbf{8.1}\\

 \hline
\end{tabular}

\label{tab:comparison_table}
\end{table}
\endgroup

\section{Conclusion}\label{sec: conclusion}

This paper augments "Hindi-English" code-Switched audio Mel spectrograms by using the proposed GAN architecture, which utilizes the $\log{F0}$ contour of the audio sample. The GAN was trained on the minority "Hindi-English" code switched class and evaluated using the Frechet Inception Distance (FID). A decrease in the FID value is observed as the training of the proposed GAN progressed, thus generating good-quality Mel spectrograms. A new dataset was created for three-class classification, taking samples from existing Indic speech corpora, which were used for spoken LID and a variety of other tasks in the speech domain. To validate the augmentation results of the generated code-switched audio samples, a three-class classifier was trained on the created dataset and the results were compared with those of a CRNN-based architecture. An increase of 4.7\% in the UAR was observed, whereas an 8.1\% increase in the F1 score values was observed compared to the baseline methods used to augment raw audio as well as spectrograms in the data. This study is a critical demonstration of the representation-learning capacity of GANs. GAN-based techniques can be utilized to address the data imbalance problem and make the models robust to changes in data. It has been observed that the effectiveness of GANs can be further increased using higher-resolution spectrograms (both on the frequency and temporal axis),techniques such as progressive growing of GANS, StyleGANs, CycleGANs,etc and using phonetic information for the conditioning of GANs. 
This technique can be further used in the ASR domain by using phonetic information to produce realistic speech utterances.

\section{Acknowledgements}
We would like to acknowledge the Samsung Prism team, Samsung R\&D, Bangalore, India., for motivating, helping, and guiding us throughout this project. This project was taken as a part of the Samsung Prism programme. The experiments are carried out under the guidance of Samsung Prism and College of Engineering, Pune.
  
\bibliographystyle{ieeetr}
\bibliography{citation}

\vspace{12pt}

\end{document}